\documentclass[sigconf]{acmart}
\usepackage{booktabs}
\usepackage{multirow}
\usepackage{pgfplots} 
\usepackage{colortbl}
\usepackage{pgf-pie} 
\usepackage{color,xcolor}
\usepackage{pifont}
\usepackage{framed}
\usepackage{algorithm}  
\usepackage{algpseudocode}  
\usepackage{amsthm,amsmath,amssymb}
\usepackage{mathrsfs}
\usepackage{listings}
\usepackage{tcolorbox}
\usepackage{scalefnt}
\newcommand{\CodeIn}[1]{{\small\texttt{#1}}}
\renewcommand\footnotetextcopyrightpermission[1]{} 
\AtBeginDocument{%
  \providecommand\BibTeX{{%
    \normalfont B\kern-0.5em{\scshape i\kern-0.25em b}\kern-0.8em\TeX}}}

\setcopyright{acmcopyright}
\copyrightyear{2022}
\acmYear{2022}
\acmDOI{10.1145/1122445.1122456}

\newcommand{\GDsmith}[0]{GDsmith}

\begin{document}

\title{\GDsmith{}: Detecting Bugs in Graph Database Engines}


\author{Wei Lin}
\affiliation{%
  \institution{Peking University}
  \city{Beijing}
  \country{China}}
  
\author{Ziyue Hua}
\affiliation{%
  \institution{Peking University}
  \city{Beijing}
  \country{China}}
  
\author{Luyao Ren}
\affiliation{%
  \institution{Peking University}
  \city{Beijing}
  \country{China}}

\author{Zongyang Li}
\affiliation{%
  \institution{Peking University}
  \city{Beijing}
  \country{China}}
  
\author{Lu Zhang}
\affiliation{%
  \institution{Peking University}
  \city{Beijing}
  \country{China}}
  
\author{Tao Xie}
\affiliation{%
  \institution{Peking University}
  \city{Beijing}
  \country{China}}
  

\begin{abstract}
Graph database engines stand out in the era of big data for their efficiency of modeling and processing linked data. There is a strong need of testing graph database engines. However, random testing, the most practical way of automated test generation, faces the challenges of semantic validity, non-empty result, and behavior diversity to detect bugs in graph database engines. To address these challenges, in this paper, we propose \GDsmith{}, the first black-box approach for testing graph database engines. It ensures that each randomly generated Cypher query satisfies the semantic requirements via skeleton generation and completion. \GDsmith{} includes our technique to increase the probability of producing Cypher queries that return non-empty results by leveraging three types of structural mutation strategies. \GDsmith{} also includes our technique to improve the behavior diversity of the generated Cypher queries by selecting property keys according to their previous frequencies when generating new queries. Our evaluation results demonstrate that \GDsmith{} is effective and efficient for automated query generation and substantially outperforms the baseline. \GDsmith{} successfully detects \textbf{27 previously unknown bugs} on the released versions of three popular open-source graph database engines and receive positive feedback from their developers.
\end{abstract}



\maketitle

\section{Introduction}

In recent years, graph database engines have been widely used in database applications from various domains, such as knowledge reasoning systems~\cite{DBLP:conf/amia/WangWLJWML20} and recommender systems~\cite{DBLP:journals/sncs/SenMGS21,DBLP:conf/uic/KonnoHBH17}. Graph database engines use graph structures for semantic queries with nodes, relationships, and properties to represent and store data~\cite{1998Artificial}. As a well-known representative of graph database engines, Neo4j~\cite{Neo4j} has long been ranked first in the DB-Engines Ranking~\cite{dbranking} (which ranks graph database engines monthly based on their popularity). Over 800 enterprise customers use Neo4j, including over 75$\%$ of Fortune 100 companies~\cite{dbranking}.
Although there is currently no query language standard for graph databases, it is generally believed that Cypher~\cite{DBLP:conf/sigmod/FrancisGGLLMPRS18,DBLP:journals/corr/abs-1802-09984}, originally contributed by Neo4j, is the most widely adopted query language specially designed for graph databases because of Neo4j's overwhelming market share~\cite{openCypher}.
As an open query language, Cypher is now used by over 10 other graph databases (e.g., RedisGraph~\cite{RedisGraph} and Memgraph~\cite{Memgraph}) and tens of thousands of developers~\cite{openCypher}.
Some graph databases that natively support other graph query languages (e.g., Gremlin~\cite{Gremlin}) are also compatible with Cypher queries via translation tools (e.g., Cypher for Gremlin~\cite{cfg}).

Detecting bugs in graph database engines is critical for two main reasons. First, the underlying code of graph database engines is of high complexity, being error-prone inevitably. For example, Neo4j version 4.3.10 has over 684K non-comment lines of Java code~\cite{neo4jcode}, and Memgraph version 2.1.1 has over 103K non-comment lines of C/C++ code~\cite{memgraphcode}. Second, the applications that interact with buggy graph database engines may exhibit unexpected failing behaviors, leading to potential risks. For example, a Neo4j bug\footnote{\url{https://github.com/neo4j/neo4j/issues/12641}} in the interpreted runtime leads to incorrect results for the \CodeIn{size} function (which is a widely used scalar function returning the number of sub-graphs matching the given pattern expression).

\begin{figure}[t]
  \centering
  \includegraphics[width=\linewidth]{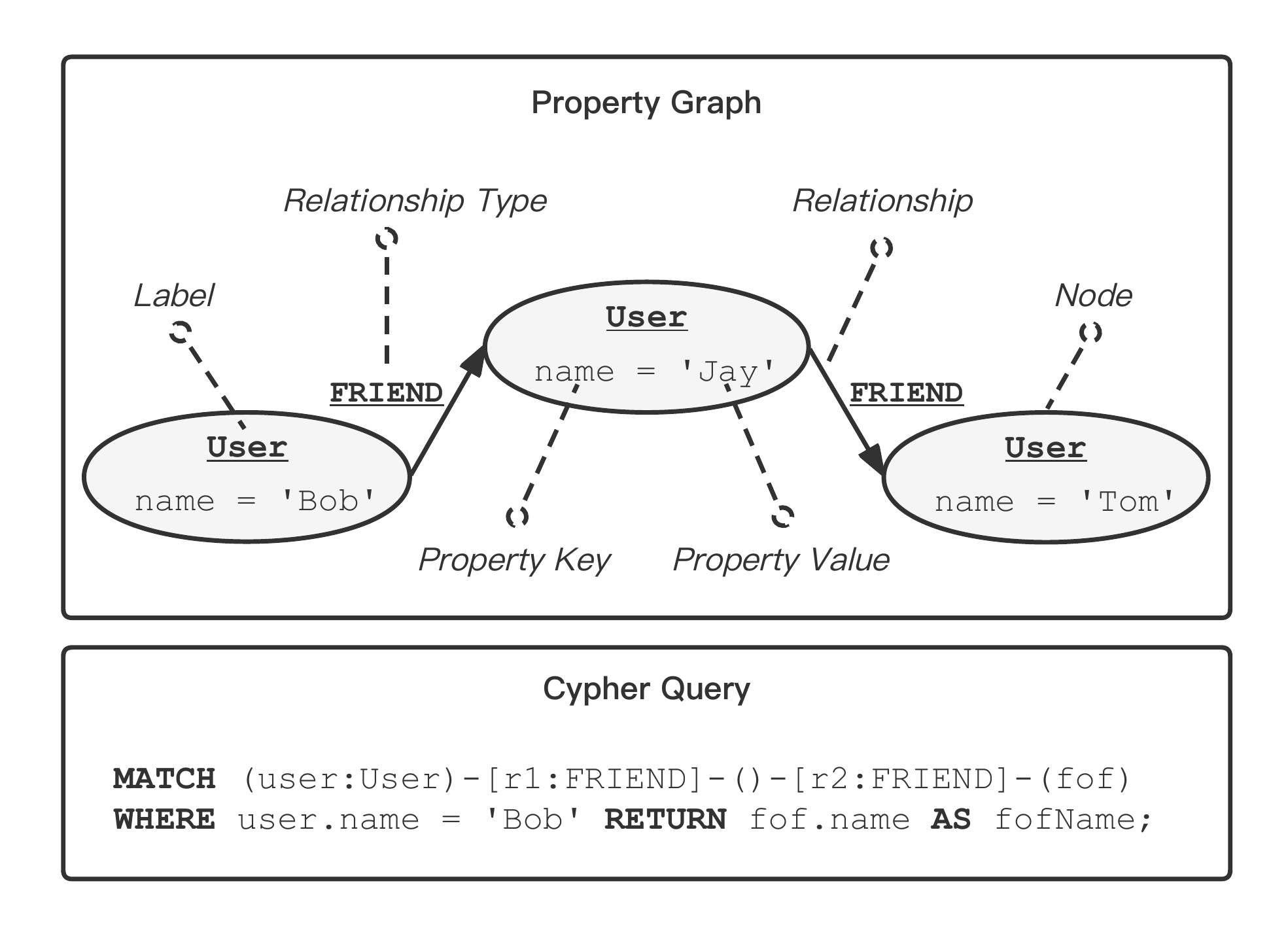}
  \caption{An example property graph (containing three \CodeIn{User} nodes and two \CodeIn{FRIEND} relationships) and a Cypher query (finding friends of friends of \CodeIn{Bob}).}
  \label{example}
\end{figure}

To detect bugs in graph database engines with Cypher support, one can adopt random testing~\cite{DBLP:conf/sigmod/GhitPRXB20,DBLP:conf/vldb/Slutz98,AFL} to generate two parts of test inputs, as shown in Figure~\ref{example}. (1) A \textbf{property graph} is a directed graph consisting of labeled entities (i.e., \emph{nodes} and \emph{relationships})\footnote{A node may be assigned with a set of unique \emph{labels}, whereas a relationship is assigned with exactly one \emph{relationship type}.} and \emph{properties} on entities~\cite{DBLP:journals/corr/abs-1710-04470}. A relationship encodes a directed connection between exactly two nodes. A property is a pair consisting of a \emph{property key} and a \emph{property value} (which is an instantiation of one of Cypher’s types such as \CodeIn{String}). (2) \textbf{Cypher queries} are built up using various clauses for querying the property graph~\cite{openCypher}. In each Cypher query, clauses are chained together, and they feed intermediate result sets between each other. For example, the matching variables from one \CodeIn{MATCH} clause will be the context that the next clause exists in. The last part of the Cypher query is one final \CodeIn{RETURN} clause (which is a period symbol at the end of a Cypher query).

However, adopting existing random testing approaches to detect bugs in graph database engines with Cypher support faces three main challenges. (1) \textbf{Semantic validity challenge.} It is challenging to randomly generate semantically valid Cypher queries.
A semantically valid Cypher query has no grammar error (e.g., using wrong keywords) and semantic error (e.g., using undefined variables).
Grammar-based testing~\cite{DBLP:conf/sigmod/GhitPRXB20,DBLP:conf/vldb/Slutz98} generates syntactically correct test inputs but wastes testing budget on generating a large number of semantically invalid test inputs. Mutation-based testing~\cite{AFL} produces new test inputs by mutating characters or bits of seed inputs. Such mutations easily lead to invalid semantics and even incorrect syntax because they do not grasp high-level concepts (e.g., scope of variables). However, only semantically valid queries are able to reach core parts of graph database engines and detect deep bugs because semantically invalid queries are filtered out by the engines' semantic check. (2) \textbf{Non-empty result challenge.} It is challenging to increase the percentage of Cypher queries returning non-empty results among all randomly generated queries. When the property graphs stored in multiple engines are the same, a bug can be detected if the same Cypher query is executed on these engines and their fetched result sets are different. Although simple queries with loose constraints easily return non-empty results~\cite{DBLP:conf/edbt/MishraK09}, randomly generated non-trivial queries are more likely to trigger bugs. However, randomly generated non-trivial queries tend to return empty results because there is no sub-graph matching the randomly generated constraints from these queries to the patterns. (3) \textbf{Behavior diversity challenge.} It is challenging to randomly generate Cypher queries with high behavior diversity. If two Cypher queries cover different property values\footnote{A covered property value refers to that it involves a query's execution and affects its result set.} in a property graph, these two Cypher queries have different behaviors. Improving the behavior diversity of randomly generated queries is desirable because some bugs can be triggered by only specific behaviors (e.g., covering specific property values). However, the existing random testing approaches lack feedback mechanisms to generate new queries covering different property values to improve the behavior diversity in a target manner.

To address the aforementioned challenges, in this paper, we propose \GDsmith{}, the first black-box approach for testing graph database engines. To address the semantic validity challenge, \GDsmith{} first randomly generates a Cypher skeleton that contains uninstantiated parts (e.g., patterns and expressions), then fills the uninstantiated parts with validly defined variables based on the semantics of each clause to obtain a Cypher query. To address the non-empty result challenge, \GDsmith{} includes our technique to increase the probability of producing Cypher queries returning non-empty results by leveraging three types of structural mutation strategies, so that the mutated queries via these strategies not only preserve semantic validity, but also return non-empty results with high probability. To address the behavior diversity challenge, \GDsmith{} dynamically selects property keys according to their previous frequencies when completing new skeletons. The insight is that high behavior diversity can be achieved with evenly distributed frequencies of property keys in Cypher queries returning non-empty results.

We implement \GDsmith{} in Java and conduct a series of evaluations to assess its effectiveness and efficiency measured with five metrics. Our evaluation results show that \GDsmith{} substantially outperforms the  baseline. \GDsmith{} achieves 100$\%$ semantic validity rate and 98$\%$ grammar coverage, whereas the baseline achieves only 16$\%$ semantic validity rate. \GDsmith{} is able to produce 69$\%$ more queries returning non-empty results than the baseline. The graph mutation score of \GDsmith{} is 4.2$\times$ higher than the graph mutation score of the baseline. \GDsmith{} detects three unique bugs on two released versions of Neo4j within 30 minutes whereas the baseline does not detect any bug.

We deploy \GDsmith{} to detect bugs in three popular open-source graph database engines (Neo4j~\cite{Neo4j}, RedisGraph~\cite{RedisGraph}, and Memgraph~\cite{Memgraph}), detecting \textbf{27 previously unknown bugs} on the released versions. Among the 27 detected bugs, 14 are confirmed by the developers of corresponding engines and 7 are already fixed. The developers of all the three graph database engines have replied that our work contributes to their development. The positive feedback from the developers also shows \GDsmith{}’s high value in practice.

This paper makes the following main contributions:
\begin{itemize}
    \item {\bfseries \GDsmith{}}, the first approach of automated test generation for detecting bugs in graph database engines, addressing three main challenges.
    \item {\bfseries Evaluations} for demonstrating \GDsmith{}'s effectiveness (substantially outperforming the baseline) and practicability (successfully detecting 27 previously unknown bugs).
\end{itemize}

The rest of the paper is organized as follows. Section~\ref{sec:background} presents the background of our work. Section~\ref{sec:approach} illustrates our \GDsmith{} approach. Section~\ref{sec:evaluation} describes our evaluations. Section~\ref{sec:work} discusses related work. Section~\ref{sec:conclusion} concludes this paper with future work.

\section{Background}
\label{sec:background}

It is known to all that graph database engines are critical software systems. They play an important role in storing and processing linked data.
In this section, we provide background information on three graph database engines with Cypher support (which are tested by \GDsmith{} in our evaluations), the Cypher language, and the historical bug statistics for these engines.

\subsection{Graph Database Engines We Test}

We select three popular open-source graph database engines as our test subjects:
\begin{enumerate}
    \item Neo4j~\cite{Neo4j} is the market leader, graph database category creator, and the most widely deployed graph data platform in the market according to the DB-Engines Ranking~\cite{dbranking}. It is a high-performance graph store with all the features expected of a mature and robust database.
    \item RedisGraph~\cite{RedisGraph} is the first queryable property graph database to use sparse matrices to represent the adjacency matrix in graphs and linear algebra to query the graph.
    \item Memgraph~\cite{Memgraph} is a streaming graph application platform and leverages an in-memory-first architecture.
\end{enumerate}

\subsection{Cypher Language}

Cypher is the most widely adopted, fully-specified, and open query language for property graph database engines~\cite{openCypher}. It is a declarative graph query language that allows for expressively and efficiently querying the graph store. Complicated database queries can easily be expressed through Cypher. Although there is no standard language for graph databases, Cypher is a key inspiration for the ISO project creating a standard graph query language (GQL) according to the official description~\cite{openCypher}. There are also existing translation tools from Cypher to other graph query languages~\cite{cfg}.

Cypher lets users simply express what data to retrieve (declarative) while the underlying engine completes the task without requiring they understand implementation details, which is in contrast to imperative languages like Java, scripting languages like Gremlin. In each Cypher query, clauses are chained together, and they feed intermediate result sets between each other. Table~\ref{tab:grammar} shows the core syntax of Cypher\footnote{The complete syntax and semantics of core Cypher can be referred to in previous work~\cite{DBLP:conf/sigmod/FrancisGGLLMPRS18,DBLP:journals/corr/abs-1802-09984}.}.
Note that \emph{a} denotes an alias, \emph{expr} denotes an expression, \emph{node\_pattern} denotes a node pattern, and \emph{rel\_pattern} denotes a relationship pattern.
The \CodeIn{MATCH} clause is used to specify the patterns the Cypher query will search for in the database. \CodeIn{MATCH} is often coupled to a \CodeIn{WHERE} part which is a sub-clause and adds restrictions to the \CodeIn{MATCH} patterns, making them more specific. A \CodeIn{MATCH} clause can occur at the beginning of the query or later, possibly after a \CodeIn{WITH} clause. The \CodeIn{WITH} clause allows query parts to be chained together, piping the results from one to be used as starting points or criteria in the next. The \CodeIn{UNWIND} clause expands a list into a sequence of records. The \CodeIn{RETURN} clause defines what to include in the query result set.
Figure~\ref{example} shows an example Cypher query that aims to find all \CodeIn{fof}'s names who are friends of friends of \CodeIn{Bob}.
In this query, \CodeIn{user:User} and \CodeIn{fof} are node patterns, \CodeIn{-[r1:FRIEND]-} and \CodeIn{-[r2:FRIEND]-} are relationship patterns, \CodeIn{user.name = 'Bob'} and \CodeIn{fof.name} are expressions, and \CodeIn{fofName} is an alias.

\begin{table}[]
\caption{Core Grammar of Cypher Queries}
\label{tab:grammar}
\resizebox{\linewidth}{!}{%
\begin{tabular}{rcl}
\hline
query                & ::=                  & \CodeIn{RETURN} ret | clause query                            \\
ret                  & ::=                  & \CodeIn{*} | \emph{expr} {[}\CodeIn{AS} \emph{a}{]} | ret\CodeIn{,} \emph{expr} {[}\CodeIn{AS} \emph{a}{]}           \\
clause               & ::=                  & {[}\CodeIn{OPTIONAL}{]} \CodeIn{MATCH} pattern\_tuple {[}\CodeIn{WHERE} \emph{expr}{]} \\
                     &                      & | \CodeIn{WITH} ret {[}\CodeIn{WHERE} \emph{expr}{]} | \CodeIn{UNWIND} \emph{expr} \CodeIn{AS} \emph{a}       \\
pattern\_tuple       & ::=                  & pattern | pattern\_tuple                             \\
pattern              & ::=                  & \emph{node\_pattern} | \emph{node\_pattern} \emph{rel\_pattern} pattern   \\ \hline
\end{tabular}%
}
\end{table}

\subsection{Historical Bug Statistics}

To gain a good understanding of bugs in graph database engines, we analyze the issues and pull requests in the official Neo4j, RedisGraph, and Memgraph GitHub repositories before December 31, 2021. We regard bug-labeled issues and pull requests as reported bugs because the official repositories of these three graph database engines happen to contain \CodeIn{bug} labels (which are used to mark possible problems). Note that such filtering is not perfect, and may overlook bugs (some platform-related issues are solved by changing graph database engine versions instead of marking these issues as bugs) and introduce redundant statistics (some bug-labeled pull requests have additional feature updates). From our final statistics, those with \CodeIn{bug} labels account for about 15$\%$ (2363 out of 15063) of all issues and pull requests.

\begin{figure}[t]
  \centering
  \includegraphics[width=\linewidth]{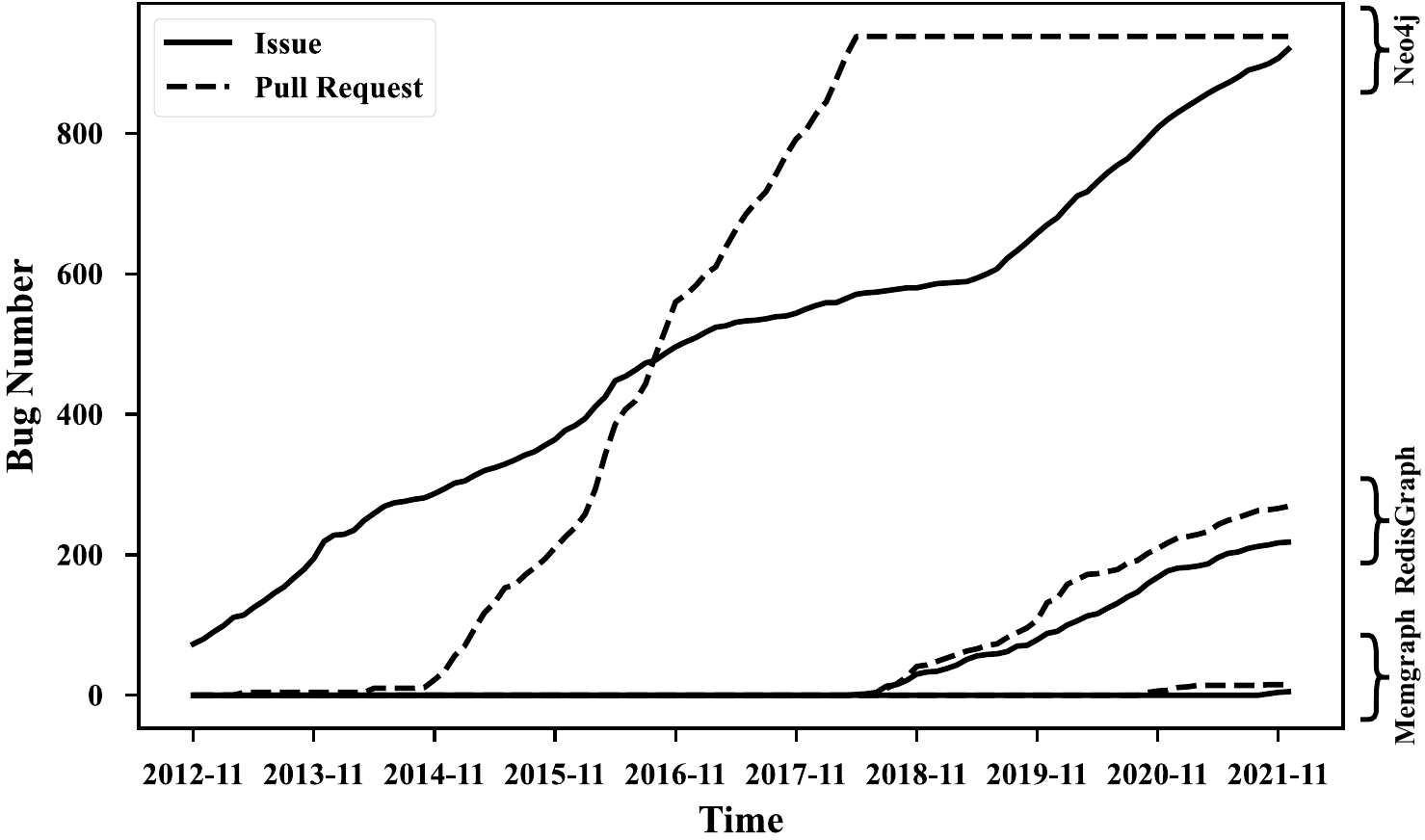}
  \caption{The overall evolution history of the bugs.}
  \label{S2RQ1}
\end{figure}

We first count the number of bug-labeled issues and pull requests in Neo4j, RedisGraph, and Memgraph by month, and plot the growth curve as shown in Figure~\ref{S2RQ1}. Neo4j, RedisGraph, and Memgraph have 920, 218, and 5 bug-labeled issues, respectively. In addition, Neo4j, RedisGraph, and Memgraph have 937, 268, and 15 bug-labeled pull requests, respectively.
The earliest issue of Neo4j is in November 2012, corresponding to Neo4j version 1.8\footnote{\url{https://neo4j.com/release-notes/page/15/}}. Since this release, the number of bug-labeled issues has increased gradually over time. The first bug-labeled pull request appears in December 2013 and since then the number of bug-labeled pull requests starts to grow. The growing trend of pull requests continues into 2018, and after that there is no new bug label added. We manually inspect all pull requests from January 2018, and they mainly contain some feature updates, with very few serious bug fixes. All of these pull requests analyzed by us are closed except that two are in the merged state and two other are still open. Since the bug-labeled issues continue to grow from the beginning of 2018 until now, we believe that all bugs raised by users during this period are continuously fixed in the official version updates.
Although RedisGraph and Memgraph are released relatively late, since their first releases the users are finding bugs in both graph database engines, and the overall number of bugs is steadily increasing. As the software functionality continues to improve, we have reasons to believe that the sizes of the bugs and the pull requests in these two databases may grow to the similar sizes as Neo4j. The growing trend of bugs in graph database engines illustrates that there is a strong need for deeply testing graph database engines.

Then we count how many bug-labeled issues contain Cypher queries. We search for the Markdown code snippets in the issue contents (only 1 out of 1143 issues does not contain any code snippet), and determine code snippets to include Cypher queries when the code snippets include specific Cypher syntax keywords such as \CodeIn{MATCH} and \CodeIn{UNWIND} (note that we do not use some other Cypher syntax keywords such as \CodeIn{AND} and \CodeIn{OR} because they may appear in code or commit logs). 
We assume that an issue containing more than two used Cypher keywords in a code snippet is considered to contain Cypher queries. This statistical mechanism may introduce some bias, as sometimes users provide detailed Cypher queries after being inquired by developers.
Neo4j, RedisGraph, and Memgraph have 358, 133, and 3 bug-labeled issues containing Cypher queries, respectively. It indicates that many bugs in graph database engines can be triggered by executing Cypher queries. Among all closed issues (1043 out of 1143), the bug living time for issues containing Cypher queries is an average of 192 days, whereas the bug living time for the other issues is an average of 298 days (36$\%$ less). We conduct further analysis in the order of bug living time and find that there is no difference at the 25$\%$ median, but it has 48$\%$ and 57$\%$ less time at the 50$\%$ and 75$\%$ median, respectively. These results indicate the potential importance of providing bug-revealing Cypher queries in the issues for speeding up solving difficult bugs during the debugging process.

\section{Approach}
\label{sec:approach}


In this paper, we propose \GDsmith{}, the first black-box approach for testing graph database engines. In particular, \GDsmith{} users specify multiple different instances of graph database engines under test. \GDsmith{} is portable and compatible without any requirement of code instrumentation, so it can also test graph database engines whose underlying code is unavailable or incomplete. Given the engines under test, \GDsmith{} automatically outputs test inputs (each including both a database state and Cypher queries) that trigger bugs. To detect whether a bug is triggered (with a test oracle), besides crashing or not, \GDsmith{} can leverage differential testing~\cite{DBLP:journals/dtj/McKeeman98} to detect bugs. For example, \GDsmith{} users can use cross-engines (i.e., comparing results fetched by different graph database engines), cross-versions (i.e., comparing results fetched by different versions of the same graph database engine), or cross-optimization (i.e., comparing results fetched by different query options on the same graph database engine).

The overall algorithm of \GDsmith{} is shown in Algorithm~\ref{overall}. In particular \GDsmith{} generates test inputs through four-step iterations. First, \GDsmith{} randomly generates a database state (Lines 2-3, shown in Section~\ref{graphgen}) and initializes \GDsmith{}'s configuration variables (Lines 4-5). Second, \GDsmith{} produces semantically valid Cypher queries via random generation or mutation (Lines 8-12, shown in Section~\ref{Skeleton and its Completion}). Third, \GDsmith{} executes generated queries on each engine instance (Lines 13-14). Fourth, \GDsmith{} retains the Cypher queries that return non-empty results (Lines 19-21, shown in Section~\ref{mutation}), and leverages the feedback of property key frequencies to generate new Cypher queries with high behavior diversity (Line 22, shown in Section~\ref{pks}). In the end, \GDsmith{} automatically outputs a bug report including the property graph (i.e., database state) and Cypher queries that trigger bugs (Lines 15-17, shown in Section~\ref{detection}). In the rest of the section, we illustrate the details of these steps. 

\begin{algorithm}[t]  
  \caption{The top-level algorithm of \GDsmith{}}  
  \label{overall}  
  \begin{algorithmic}[1]
    \Require  
    $D_a$: one instance of a graph database engine;
    \Require  
    $D_b$: the other instance of a graph database engine;
    \Require
    $N_g$: the maximum number of newly generated queries;
    \Require
    $N_m$: the maximum number of mutated queries;
    \Ensure  
    $B$: bug reports.
    \While {the timeout does not exceed}
        \State $S \gets$ \textsc{GeneratePropertyGraphSchema}($D_a, D_b$)
        \State $G \gets$ \textsc{GeneratePropertyGraph}($S, D_a, D_b$)
        \State $Pool \gets \emptyset$
        \State $F \gets$ Set all frequencies to zero.
        \State $i \gets 0$
        \While {$i < N_g + N_m$}
            \If {$i < N_g$}
                \State $Q \gets$ \textsc{GenerateRandomQuery}($S, F$)
            \Else
                \State $Q \gets$ \textsc{MutatingExistingQuery}($S, F, Pool$)
            \EndIf
            \State $R_a \gets$ \textsc{ExecuteCypherQuery}($Q, D_a$)
            \State $R_b \gets$ \textsc{ExecuteCypherQuery}($Q, D_b$)
            \If {a bug is detected by comparing $R_a$ with $R_b$}
                \State $B \gets$ Generate the bug report including $G$ and $Q$.
                \State \Return{$B$}
            \EndIf
            \If {$R_a$ or $R_b$ is non-empty}
                \State $Pool \gets$ Retain $Q$ into $Pool$
            \EndIf
            \State $F \gets$ Update the frequencies of property keys in $Q$.
            \State $i \gets i + 1$
        \EndWhile
    \EndWhile
  \end{algorithmic}  
\end{algorithm}

\subsection{Schema and Graph Generation}
\label{graphgen}

In each iteration, \GDsmith{} first randomly generates the property graph schema. A property graph schema defines the labels, the relationship types, and the properties on them~\cite{DBLP:journals/corr/abs-1710-04470}. \GDsmith{} generates a finite set of labels. For each label it determines a unique name and a set of properties. \GDsmith{} also generates a finite set of relationship types. For each relationship type it determines a unique name and a set of pairs of labels for which the relationship type is applicable. \GDsmith{} generates a unique name and a data type for each property. Generating a property graph schema is necessary because it helps guide the valid generation of a property graph and Cypher queries. Note that some graph database engines are schema-free so there is no need to execute statements to create the property graph schema in advance, others are schema-based so the property graph schema needs to be created explicitly.

\GDsmith{} then randomly generates a property graph and insert it into each engine instance under test. \GDsmith{} specifies the number of nodes and relationships in the property graph, and then sequentially generates property values in each node or relationship based on the generated property graph schema. After graph generation, \GDsmith{} also creates the same random indexes for all engine instances under test.

\subsection{Skeleton and its Completion}
\label{Skeleton and its Completion}

To address the semantic validity challenge, \GDsmith{} includes a skeleton-based completion technique to generate each Cypher query. We refer to each clause sequence with uninstantiated parts (e.g., patterns and expressions) as a \emph{skeleton}, and an uninstantiated part is denoted as the $\square$ symbol. Specifically, the skeletons generated by \GDsmith{} are the languages that are produced by the grammar of Table~\ref{tab:ske}.

For example, the skeleton corresponding to the Cypher query in Figure~\ref{example} is ``\CodeIn{\textbf{MATCH} $\square$ \textbf{WHERE} $\square$ \textbf{RETURN} $\square$}''.

\begin{table}[]
\caption{Core Grammar of Cypher Skeletons}
\label{tab:ske}
\begin{tabular}{rcl}
\hline
skeleton                & ::=                  & \CodeIn{RETURN} ret | clause skeleton                            \\
ret                  & ::=                  & \CodeIn{*} | $\square$           \\
clause               & ::=                  & {[}\CodeIn{OPTIONAL}{]} \CodeIn{MATCH} $\square$ {[}\CodeIn{WHERE} $\square${]} \\
                     &                      & | \CodeIn{WITH} ret {[}\CodeIn{WHERE} $\square${]} | \CodeIn{UNWIND} $\square$   \\ \hline
\end{tabular}
\end{table}

\begin{algorithm}[t]  
  \caption{The \textsc{GenerateRandomQuery} function}  
  \label{GenerateRandomIR}  
  \begin{algorithmic}[1]

    \Require  
    $S$: the property graph schema;
    \Require  
    $F$: the frequency list for property keys;
    \Ensure
    $Q$: the newly generated Cypher query.
    \State $N_o \gets$ Randomly determine the number of clauses.
    \State $Skeleton \gets$ \textsc{GenerateSkeleton}($N_o$)
    \State $Q \gets$ $\emptyset$
    \State $Context \gets$ $\emptyset$
    \State $i \gets 0$
    \While {$i < N_o$}
        \State $O_s \gets$ Get the $i$-th clause in $Skeleton$.
        \State $O_c \gets$ \textsc{FillUninstantiatedPart}($O_s, S, F, Context$)
        \State $Q \gets$ Append $O_c$ to $Q$.
        \State $Context \gets$ \textsc{CalculateNewContext}($O_c$, $Context$)
        \State $i \gets i + 1$
    \EndWhile
    \State \Return{$Q$}
  \end{algorithmic}  
\end{algorithm}

The random generation of each Cypher query is divided into two steps as shown in Algorithm~\ref{GenerateRandomIR}. \GDsmith{} firstly randomly generates a Cypher skeleton (Lines 1-2). A skeleton is an incomplete Cypher query conducted by a clause sequence with uninstantiated parts (i.e., ``$\square$''). All the skeletons can be considered as a language $\mathcal{L}_{skeleton}$ generated by the grammar shown in Table~\ref{tab:ske}. $\mathcal{L}_{skeleton}$ is context-free and thus the generation of a skeleton requires rarely additional semantic information\footnote{Some graph database engines do not support Cypher queries where a \texttt{OPTIONAL MATCH} clause is followed by a \texttt{MATCH} clause. \GDsmith{} filters out such skeletons when testing these graph database engines.}. Therefore, \GDsmith{} randomly generates a clause sequence with uninstantiated parts appended by a \CodeIn{RETURN} clause. 

When filling the uninstantiated parts in the skeleton, \GDsmith{} calculates and maintains each clause's context.
A context includes the following two kinds of information:
\begin{enumerate}
    \item  {\bfseries Local environment}: the local environment of each clause is a set of variables available in the next clause. A variable is an identifier that represents a vector of data defined in a pattern or an alias definition (e.g., literals or entities). For a \CodeIn{MATCH} clause, \GDsmith{} adds the variables in the previous clause's local environment and the newly defined entity variables in node patterns and relationship patterns into its local environment. For an \CodeIn{UNWIND} clause, \GDsmith{} adds the variables in the previous clause's local environment and the newly defined variable (i.e., alias) into its local environment. For a \CodeIn{WITH} or \CodeIn{RETURN} clause, \GDsmith{} adds the variables defined in expressions and aliases into its local environment. Take the Cypher query in Figure~\ref{example} as an example, the local environment of the \CodeIn{MATCH} clause is \CodeIn{\{user, r1, r2, fof\}}.

    \item {\bfseries Fact}: the fact of each clause is the information collected from the variable definition. A fact contains the data type of each variable. \GDsmith{} calculates a variable's data type by analyzing its definition. In detail, if a variable is defined in \CodeIn{node\_pattern} or \CodeIn{rel\_pattern}, it is an entity. Otherwise, the variable is defined as an alias of an expression. \GDsmith{} can then calculate the variable's data type by analyzing the expression's output type. If a variable's data type is a node, \GDsmith{} also maintains the label constraints of this variable in the fact. If a variable's data type is a relationship, \GDsmith{} also maintains the type constraints of this variable in the fact. Take the Cypher query in Figure~\ref{example} as an example, the label/type constraints in the fact of the \CodeIn{MATCH} clause is \CodeIn{\{label(user) == [User], type(r1) == [FRIEND], type(r2) == [FRIEND], label(fof) == [User]\}}.
\end{enumerate}

Based on rich semantic information provided by the property graph schema and the context of each clause, \GDsmith{} is able to make the generated uninstantiated parts satisfy the following three semantic requirements:
\begin{enumerate}
    \item {\bfseries Variable scope correctness}: a variable can be referenced in a clause only when this variable is contained in the local environment of the clause.
    \item {\bfseries Operand type safety}: the operand type of each expression must satisfy the type requirement of the expression. In addition, the combination of operands must assure the output type of the expression is inferable at runtime.
    \item {\bfseries Property key safety}: all property keys taken from entities must exist in the property graph schema at runtime.
\end{enumerate}

In detail, to ensure the three semantic requirements, \GDsmith{} uses the strategies to fill uninstantiated parts as follows (Line 8).
\begin{itemize}
    \item If the current clause requires a pattern tuple, \GDsmith{} dynamically generates a series of patterns for it. Each pattern is constructed by a sub-graph described by entity variables. \GDsmith{} firstly generates a shape of the sub-graph (e.g., a node with a relationship points to another node). For each variable required in the shape, \GDsmith{} either chooses a defined variable in the previous clause's local environment, or defines a new variable for it. Then, \GDsmith{} randomly adds new labels or types to these variables. Finally, \GDsmith{} randomly adds properties to these variables according to the property graph schema and the related fact.
    \item If the current clause requires an expression with a specific data type, \GDsmith{} uses a top-down strategy to generate the expression. An expression is a tree structure where each tree node is a sub-expression. For each operand that requires a sub-expression, \GDsmith{} randomly chooses an expression that satisfies the type requirement and generates the whole sub-tree recursively. For each operand that requires a literal value, \GDsmith{} randomly generates a literal value of the required type. For each operand that requires a variable reference, \GDsmith{} searches for a variable with required type from the context. If no suitable variable is found, \GDsmith{} discards the current sub-tree and re-generates it. The expression generation process is not stopped until all operands of current expression tree are filled or the maximum depth is reached.
\end{itemize}

After filling all uninstantiated parts in a clause, \GDsmith{} will calculate the context of this clause according to its definition to guide the completion of the next clause (Line 10).

\subsection{Structural Mutation Strategies}
\label{mutation}

To address the non-empty result challenge, \GDsmith{} leverages three structural mutation strategies to increase the probability of producing Cypher queries returning non-empty results. \GDsmith{} retains each generated Cypher query that returns non-empty results into a pool. These retained queries are used for producing new queries via structural mutation strategies in the next stage. This feedback increases the probability that a newly mutated query returns non-empty results. In addition, \GDsmith{} requires the number of clauses in each retained query to be within an appropriate range. A small number of clauses indicate that the Cypher query may be simple and trivial, reducing the diversity of subsequent mutations to produce new queries. A large number of clauses indicate that the Cypher query may be complicated, reducing the efficiency of bug detection because its mutants may take longer time to be executed.

\begin{algorithm}[t]  
      \caption{The \textsc{MutatingExistingQuery} function}  
  \label{MutatingExistingIR}  
  \begin{algorithmic}[1]
    \Require  
    $S$: the property graph schema;
    \Require  
    $F$: the frequency list for property keys;
    \Require  
    $Pool$: the retained query pool;
    \Ensure
    $Q$: the newly mutated Cypher query.
    \State $Q_e \gets$ Randomly select a Cypher query from $Pool$.
    \State $Q \gets$ $\emptyset$
    \State $T \gets$ Randomly choose a mutation strategy.
    \If {$T = ``DelayReturn"$}
        \State $Skeleton \gets$ Change the \CodeIn{RETURN} clause in $Q_e$ to \CodeIn{WITH}.
        \State $Q \gets$ Extend and complete $Skeleton$ based on $S$ and $F$.
    \ElsIf {$T = ``AdvanceReturn"$}
        \State $Skeleton \gets$ Change any \CodeIn{WITH} clause in $Q_e$ to \CodeIn{RETURN}.
        \State $Q \gets$ Discard the followed clauses in $Skeleton$.
    \ElsIf {$T = ``RemoveCondition"$}
        \State $Q \gets$ Randomly remove \CodeIn{WHERE} conditions in $Q_e$.
    \EndIf
    \State \Return{$Q$}
  \end{algorithmic}  
\end{algorithm}

\GDsmith{} uses three types of mutation strategies to derive a new Cypher query from an existing one, as shown in Algorithm~\ref{MutatingExistingIR}. For each call, this algorithm randomly applies one mutation strategy to the original Cypher query. The detailed strategies are as follows.

(1) \textbf{The \CodeIn{DelayReturn} strategy} (Line 4) firstly change the \CodeIn{RETURN} clause at the end of the clause sequence to a \CodeIn{WITH} clause (Line 5). Then, it generates a new clause sequence and appends it at the end of the sequence to form a completed Cypher query (Line 6). It uses the same strategy in Algorithm~\ref{GenerateRandomIR} to complete uninstantiated parts except that it takes the context at the end of original sequence as the initial context.

For example, the selected query $Q_e$ is 
\begin{framed}
\noindent \texttt{\textbf{MATCH} (n) \textbf{RETURN} n.k AS a;}
\end{framed}

After executing Lines 5-6, a new query $Q$ is generated.
\begin{framed}
\noindent \texttt{\textbf{MATCH} (n) \textbf{WITH} n.k AS a \textbf{RETURN} COUNT(a);}
\end{framed}

The \CodeIn{DelayReturn} strategy heuristically makes a mutated query more likely to return non-empty results than randomly generating a new query with the same query length\footnote{Query length refers to the number of clauses in the Cypher query (sub-clauses such as \texttt{WHERE} and \texttt{ORDER BY} are not counted).}.

(2) \textbf{The \CodeIn{AdvanceReturn} strategy} (Line 7) randomly chooses a \CodeIn{WITH} clause in the original Cypher query and translates it to a \CodeIn{RETURN} clause (Line 8). It then discards all the clauses following the newly created \CodeIn{RETURN} clause (Line 9). When replacing a \CodeIn{WITH} clause with \CodeIn{RETURN}, if the \CodeIn{WITH} clause contains a \CodeIn{WHERE} sub-clause, it will be removed.

For example, the selected query $Q_e$ is 
\begin{framed}
\noindent \texttt{\textbf{MATCH} (n)-->(m) \textbf{WITH} m.k AS a \textbf{RETURN} MAX(a);}
\end{framed}

After executing Lines 8-9, a new query $Q$ is generated.
\begin{framed}
\noindent \texttt{\textbf{MATCH} (n)-->(m) \textbf{RETURN} m.k AS a;}
\end{framed}

The \CodeIn{AdvanceReturn} strategy makes a mutated query return the intermediate result set of the corresponding original query, greatly increasing the probability of fetching non-empty results.

(3) \textbf{The \CodeIn{RemoveCondition} strategy} (Line 10) randomly removes \CodeIn{WHERE} conditions from \CodeIn{MATCH}, \CodeIn{OPTIONAL MATCH}, or \CodeIn{WITH} clauses (Line 11), by nature increasing the probability of fetching non-empty results.

For example, the selected query $Q_e$ is 
\begin{framed}
\noindent \texttt{\textbf{MATCH} (n)--(m) \textbf{WHERE} m.k > 0 \textbf{RETURN} n.k;}
\end{framed}

After executing Line 11, a new query $Q$ is generated.
\begin{framed}
\noindent \texttt{\textbf{MATCH} (n)--(m) \textbf{RETURN} n.k;}
\end{framed}

\subsection{Property Key Selection}
\label{pks}

To address the behavior diversity challenge, \GDsmith{} dynamically selects property keys according to their previous frequencies when completing new skeletons. \GDsmith{} maintains a frequency list for property keys as feedback. In each iteration after generating a property graph schema, all frequencies of property keys are set to zero. When a new query contains property keys in its clauses and returns non-empty results, the frequency list is automatically updated. The frequency list for property keys guides the selection of available property keys when filling uninstantiated parts of each incomplete Cypher skeleton. The formula of each property key’s selection probability is as follows, where $P_i$ is the selection probability of the $i$-th available property key, $F_i$ is the frequency of the $i$-th available property key, and $N$ is the total number of available property keys.

\begin{equation}\label{eq}
    P_i=\frac{\sum_{j=1}^NF_j -F_i}{(N-1)\sum_{j=1}^NF_j}
\end{equation}

This formula indicates that when specifying property keys from the list of all available property keys to fill uninstantiated parts, \GDsmith{} preferentially selects the property keys with low frequencies, so as to average the occurrence possibility of each property key in a set of Cypher queries returning non-empty results. The purpose is to cover diverse property values, thereby improving the behavior diversity of randomly generated queries.

Note that traditional feedback-guided random testing approaches usually use code coverage as feedback, which however does not apply to testing graph database engines for two main reasons. First, the code coverage appears to be low even if plenty of diverse queries are executed~\cite{DBLP:journals/pvldb/JungHAKK19}. It is reasonable because we concern about testing only the engine components related to data-centric Cypher queries. Other components such as interactive consoles are not the subjects for \GDsmith{}. Second, We expect \GDsmith{} to be a black-box approach to strengthen its compatibility. If the source code of graph database engines is unavailable or with language mixture, it is hard to conduct code instrumentation to obtain the runtime information such as code coverage.

\subsection{Bug Detection}
\label{detection}

By comparing the returned results of the same Cypher query on different instances of graph database engines, \GDsmith{} is able to detect two main types of bugs. (1) An {\bfseries error bug} is triggered if a syntactically correct and semantically valid query cannot be successfully executed (e.g., throwing an exception). Such bugs prevent users from obtaining expected results, and in more serious cases, cause the graph database engine to crash and lose connection to the database application. (2) A {\bfseries wrong-result bug} is triggered if a syntactically correct and semantically valid query is executed successfully but returns incorrect results. Such bugs are more dangerous in that users mistakenly believe that the Cypher query returns correct results and have wrong expectations about the behavior, leading to potential risks.

Note that when \GDsmith{} aims to detect wrong-result bugs, there are several constraints for each Cypher query. First, the Cypher query containing non-deterministic sub-clauses may result in false alarms. For example, the result set will get trimmed from the top by using the \CodeIn{SKIP} sub-clause. However, without an \CodeIn{ORDER BY} sub-clause, the records are randomly selected because no guarantees are made on the order of the result~\cite{openCypher}. To avoid such false positives, \GDsmith{} uses deterministic clauses and routines for query generation. Second, undefined behaviors make it hard to determine whether inconsistent results are bugs or just different implementations. For example, integer overflows and divisions by zero are handled differently by different engines (e.g., returning \CodeIn{NaN} or throwing exceptions). \GDsmith{} will not generate the Cypher queries that may execute such undefined behaviors. Third, for convenient and efficient comparison, the Cypher query should return only a few specific expressions instead of a large number of entities. For example, when executing a Cypher query whose \CodeIn{RETURN} clause contains the \CodeIn{$\star$} symbol on each graph database engine, parsing and comparing result sets including all nodes and relationships from different clients are time-consuming because an entity may consist of plenty of property values.

\section{Evaluations}
\label{sec:evaluation}

In our evaluations, we address the following four research questions (RQs):
\begin{itemize}
    \item {\bfseries RQ1:} How effective are the Cypher queries generated by \GDsmith{}?
    \item {\bfseries RQ2:} How much do different techniques contribute to the overall effectiveness and efficiency of \GDsmith{}?
    \item {\bfseries RQ3:} How much does \GDsmith{} outperform the baseline?
    \item {\bfseries RQ4:} How practicably does \GDsmith{} detect real-world bugs in popular graph database engines?
\end{itemize}

\subsection{Evaluation Setup}

\subsubsection{Subjects}

We opt for three popular real-world graph database engines as our test subjects. We test the released versions of Neo4j Community Edition from 3.5 to 4.4, the released versions of RedisGraph from 2.4 to 2.8, and the released version 2.1.1 of Memgraph Community Edition. All of these subjects are downloaded from their official repositories without any underlying modification.

\subsubsection{Implementation}

We implement the \GDsmith{} prototype with over 12K non-comment lines of Java code. Its framework is derived from SQLancer~\cite{SQLancer} (which is a tool to automatically test relational database engines). \GDsmith{} uses Neo4j Java Driver 4.1.1 to connect and interact with Neo4j and Memgraph, and uses JRedisGraph 2.5.1 to connect and interact with RedisGraph. Some graph database engines implement only a subset of the Cypher language. When conducting cross-engines differential testing, all Cypher queries generated by \GDsmith{} do not contain any Cypher feature that is unsupported by either graph database engine. All evaluations are conducted on a Windows 11 laptop with Intel i7-8565U CPU and 16 GB of memory.

\subsubsection{Baseline}

Note that there is no applicable baseline to which we can compare our work because no existing work focuses on detecting bugs in graph database engines and \GDsmith{} is the first approach for it. Therefore, we design the approach named \GDsmith{}$_{!a}$ that only guarantees the syntactic correctness of generated queries as the baseline. To produce a query, \GDsmith{}$_{!a}$ randomly generates a skeleton and then completes it with random parameters (semantic requirements are not considered). It essentially falls under grammar-based random testing.

\subsubsection{Metrics}

To measure the effectiveness and efficiency of \GDsmith{}, we design the following five types of metrics:
\begin{enumerate}
    \item The {\bfseries semantic validity rate} is defined as the percentage of semantically valid queries among all generated Cypher queries. High semantic validity rate indicates that the generated queries can reach core parts of graph database engines.
    \item The {\bfseries core grammar coverage} is defined as the percentage of covered productions~\cite{DBLP:journals/computing/Burkhardt67,Paul1972A} of semantically valid Cypher queries among all productions in the core Cypher syntax~\cite{DBLP:conf/sigmod/FrancisGGLLMPRS18,DBLP:journals/corr/abs-1802-09984}. High semantic validity rate indicates that the generated queries can cover diverse Cypher features.
    \item The {\bfseries non-empty result rate} is defined as the percentage of queries returning non-empty results among all generated Cypher queries. High non-empty result rate indicates that the generated queries are meaningful towards differential testing, especially for detecting wrong-result bugs.
    \item The {\bfseries graph mutation score} is defined as the number of different killings of a set of graph mutants\footnote{A graph mutant is generated by arbitrarily removing one property from the original property graph. A graph mutant is killed as long as one Cypher query returns different results from this graph mutant and the original graph.} by generated Cypher queries. Take three queries and three graph mutants as an example, if the first query kills the first and the second graph mutants, whereas both the second and the third queries kill the first and the third graph mutants, the graph mutation score of these three queries is two. High graph mutation score indicates that the generated queries have high behavior diversity.
    \item The {\bfseries bug detection efficiency} is defined as the number of unique bugs found within the same period. High bug detection efficiency indicates that the generated queries are able to detect bugs at a high speed.
\end{enumerate}

\subsection{RQ1/RQ2: Effectiveness and Efficiency}

To assess the effectiveness and efficiency of \GDsmith{}, we design and implement two variants of \GDsmith{} for our evaluations. One variant named \GDsmith{}$_{!m}$ disables the structural mutation strategies, and all generated queries are newly generated. The other variant named \GDsmith{}$_{!f}$ disables the feedback of property key frequencies, and all property keys are randomly selected to complete new skeletons. We conduct all the evaluations in this subsection on Neo4j because Neo4j implements the most complete Cypher semantics.

\subsubsection{Semantic Validity Rate} 

We run the three approaches (including \GDsmith{} and two variants) five times in turn. In each run, 5000 queries are generated. We count the number of Cypher queries that do not throw any semantics-related exception during execution. 

The evaluation results show that the semantic validity rates of the three approaches are all 100$\%$, demonstrating the effectiveness of our designed techniques of skeleton completion to satisfy semantic requirements. The results also confirm that \GDsmith{} is able to reach core parts of graph database engines.

\subsubsection{Core Grammar Coverage}

For each of the three approaches, we produce 5000 Cypher queries and leverage Lark~\cite{Lark} (which is a parsing toolkit for context-free languages) to parse each query and calculate the core grammar coverage.

The evaluation results show that the core grammar coverage of the three approaches all converges to 98$\%$. Only three out of 122 productions are not covered in the core Cypher syntax because we do not implement some Cypher features such as \CodeIn{UNION} clauses in our prototype yet. The results confirm that \GDsmith{} does not lose any form of possible clause sequences and parameters within the scope of our implementation.

\subsubsection{Non-empty Result Rate} 

We run the three approaches ten times in turn and in each run 5000 queries are randomly generated. We limit the maximum number of entities in each property graph to 10 and the minimum length of each Cypher query to 3 because smaller graph and longer query make it harder to fetch non-empty results.

\begin{figure}[t]
  \centering
  \includegraphics[width=0.5\textwidth]{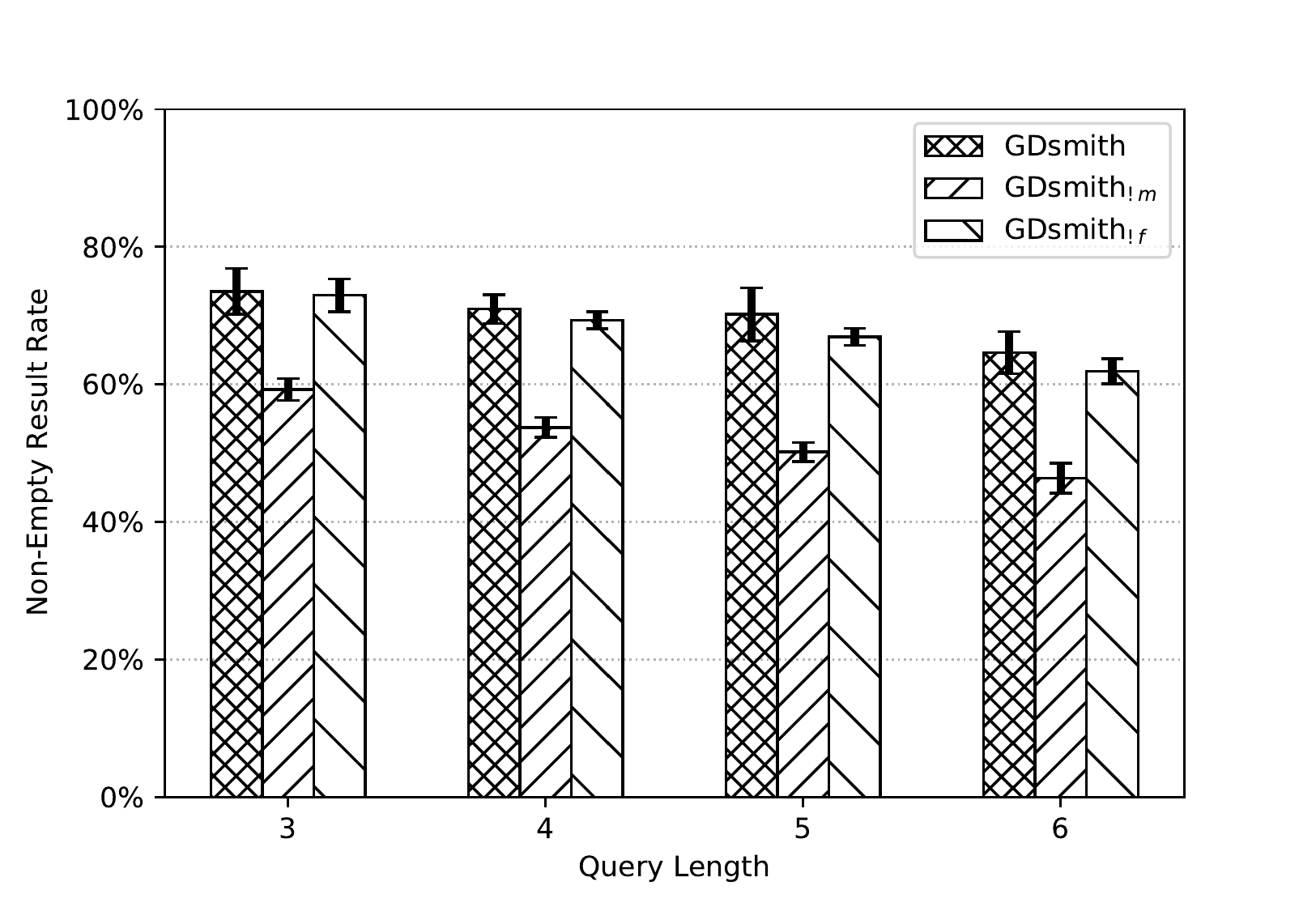}
  \caption{The non-empty result rates of the three approaches for different query lengths.}
  \label{rq3a}
\end{figure}

Figure~\ref{rq3a} shows the non-empty result rates of the three approaches for different query lengths. The evaluation results show that among all Cypher queries with lengths 3, 4, 5, and 6 generated by \GDsmith{}, the queries returning non-empty results account for 74$\%$, 71$\%$, 70$\%$, and 65$\%$, respectively. \GDsmith{} is able to produce 24$\%$ to 40$\%$ more Cypher queries returning non-empty results than \GDsmith{}$_{!m}$, and produce only 1$\%$ to 5$\%$ more Cypher queries returning non-empty results than \GDsmith{}$_{!f}$. The results illustrate that the structural mutation strategies contributes the most to increasing the non-empty result rate.

\subsubsection{Graph Mutation Score}

In this evaluation, we limit the minimum number of entities in each property graph to 20 and the maximum number of Cypher queries to 800 because larger graphs and fewer queries make it harder to cover more property values.

The evaluation results show that when generating 50 graph mutants, the graph mutation score of \GDsmith{} reaches 26, which is 10$\%$ lower than the graph mutation score of \GDsmith{}$_{!m}$ but 73$\%$ higher than the graph mutation score of \GDsmith{}$_{!f}$. The results illustrate that the feedback of property key frequencies contributes to improving behavior diversity. But to some extent, the structural mutation strategies sacrifice little behavior diversity of the mutated queries to raise the probability of returning non-empty results.

\subsubsection{Bug Detection Efficiency}

We next examine the efficiency of \GDsmith{} in detecting bugs. We also investigate whether the structural mutation strategies and the feedback of property key frequencies increase the probability of bug detection. In this evaluation, we run each of the four approaches on the two instances of Neo4j version 4.3.10 and Neo4j version 4.2.15, respectively. We configure each run to a timeout of 30 minutes, because in practice \GDsmith{} spends less than 30 minutes finding each bug in real-world graph database engines.

Within 30 minutes, \GDsmith{} generates six bug reports (reduces to two unique wrong-result bugs and one error bug), \GDsmith{}$_{!m}$ generates two bug reports (reduces to only one error bug), and \GDsmith{}$_{!f}$ generates three bug reports (reduces to one wrong-result bug and one error bug). The evaluation results show that both the structural mutation strategies and the feedback of property key frequencies have their impacts on helping \GDsmith{} detect more wrong-result bugs in graph database engines.

\subsection{RQ3: Comparison with Baseline}

We compare \GDsmith{} with the baseline (i.e., \GDsmith{}$_{!a}$) measured with the aforementioned metrics. (1) After producing 5000 queries, the semantic validity rate of the baseline reaches only 16$\%$ but its core grammar coverage also converges to 98$\%$. (2) Among all queries with lengths 3, 4, 5, and 6 generated by the baseline, the queries returning non-empty results account for 57$\%$, 45$\%$, 39$\%$, and 25$\%$, respectively. (3) The graph mutation score of the baseline reaches only 5. (4) Within 30 minutes, the baseline does not generate any bug report.

The results illustrate that the baseline generates many more semantically invalid queries than \GDsmith{}. It lacks any feedback guidance to produce high-quality Cypher queries for testing. Therefore, \GDsmith{} substantially outperforms the baseline with the high effectiveness and efficiency of detecting bugs in graph database engines.

\subsection{RQ4: Practicability}

\subsubsection{Study Findings}

Table~\ref{bug} shows, for each subject, the number of error bugs and wrong-result bugs detected in the second and third columns, respectively, and the number of fixed bugs, confirmed but unfixed bugs, and reported but unconfirmed bugs in the fourth, fifth, and sixth columns, respectively. \GDsmith{} detects 27 previously unknown bugs in total. Note that all these bugs are unique and detected on the released versions of the engines under test. Among the 27 detected bugs, 14 are confirmed by developers of corresponding engines and 7 are already fixed. In detail, \GDsmith{} detects 17 bugs by the cross-engine oracle, and detects 18 bugs by the cross-version oracle (some are overlapping with the previous oracle). \GDsmith{} does not use the cross-optimization oracle because some engines do not provide optimization options or provide optimization options in only their enterprise versions (not free). But we are confident that \GDsmith{} can use this oracle to detect bugs in the enterprise versions.

\begin{table}[t]
\caption{Bug Detection Results of \GDsmith{}}
\label{bug}
\begin{tabular}{l|rr|rrr}
\hline
\multicolumn{1}{c|}{\textbf{Subject}} & \multicolumn{1}{c}{\textbf{Err}} & \multicolumn{1}{c|}{\textbf{WR}} & \multicolumn{1}{c}{\textbf{Fixed}} & \multicolumn{1}{l}{\textbf{Confirmed}} & \multicolumn{1}{c}{\textbf{Reported}} \\ \hline
Neo4j                                 & 7                                & 8                                & 5                                  & 1                                     & 9                                     \\
RedisGraph                            & 2                                & 5                                & 2                                  & 5                                      & 0                                     \\
Memgraph                              & 0                                & 5                                & 0                                  & 1                                      & 4                                     \\ \hline
SUM                                   & 9                                & 18                               & 7                                  & 7                                     & 13                                     \\ \hline
\end{tabular}
\end{table}

\subsubsection{Selected Bugs}

Next, we show a selection of confirmed bugs found by \GDsmith{} to give an intuition on what kinds of bugs in the three graph database engines it can detect. For brevity, we show only reduced Cypher queries that demonstrate the underlying core problem, rather than the original queries and property graphs that found the bugs.

(1) The following Cypher query triggers an error bug in Neo4j version 4.3.6 (which is de facto the first bug found by \GDsmith{}). The root cause is that the clause sequence ``\CodeIn{OPTIONAL MATCH (n) MATCH (n) OPTIONAL MATCH (n)}'' causes an incorrect query plan due to \CodeIn{OPTIONAL MATCH} issues. We believe that this bug is non-trivial, since it demonstrates that even mature graph database engines are prone to such simple query.
    
\begin{framed}
\noindent \texttt{\textbf{OPTIONAL MATCH} (n) \textbf{MATCH} (n) \textbf{OPTIONAL MATCH} (n) \textbf{RETURN} n;}
\end{framed}

(2) The following graph and Cypher query triggers a wrong-result bug in Neo4j version 4.3.10. The returned results are not sorted as expected. The root cause is that in the planning process of Neo4j, a buggy optimization of the plan is conducted which ends up destroying the sort. We believe that this bug is non-trivial, since it demonstrates that wrong query optimization may trigger bugs.
    
\begin{framed}
\noindent \texttt{\textbf{CREATE} (n0), (n1); \\ \textbf{MATCH} (n0) \textbf{UNWIND} [0, 1] \textbf{AS} a \textbf{OPTIONAL MATCH} (n0), (n1) \textbf{RETURN} a \textbf{ORDER BY} a;}
\end{framed}

(3) The following graph and Cypher query triggers a wrong-result bug in RedisGraph version 2.4.11. The Cypher query should return 1 row but incorrectly returns an empty result set. The root cause is that the clause sequence ``\CodeIn{MATCH (n) OPTIONAL MATCH (n) WHERE (arbitrary branches)}'' mistakenly causes deletion of \CodeIn{n}. We believe that this bug is non-trivial, since it demonstrates that a \CodeIn{WHERE} sub-clause can incorrectly influence a query’s result set.
    
\begin{framed}
\noindent \texttt{\textbf{CREATE} (n); \\ \textbf{MATCH} (n) \textbf{OPTIONAL MATCH} (n) \textbf{WHERE} true \textbf{RETURN} n;}
\end{framed}

(4) The following graph and Cypher query triggers a wrong-result bug in Memgraph version 2.1.1. The Cypher query should return 1 row but incorrectly returns an empty result set. The root cause is that the implementation of Memgraph is not consistent with the standard semantics of Cypher language. The official documentation (i.e., Cypher Query Language Reference (Version 9)~\cite{openCypher}) states that \CodeIn{count(null)} returns 0. We believe that this bug is non-trivial, since it demonstrates that \GDsmith{} is able to detect such semantic inconsistency behaviors.
    
\begin{framed}
\noindent \texttt{\textbf{CREATE} (n); \\ \textbf{MATCH} (n:L) \textbf{RETURN} count(n) > 0 \textbf{AS} a;}
\end{framed}

\subsubsection{Developers’ Feedback}

After we report our detected bugs to the corresponding communities, developers of the three graph database engines give responses for most of the reported bugs, all with positive feedback. One of Neo4j core developers replies with the following message for bugs reported by us: ``\emph{Thank you for raising with us in the first place ... The queries all look interesting ...}'' He also said: ``\emph{We are impressed by and very curious about the query generator you have come up with. We have been paying particular attention to the issues you have raised on our project.}'' One of RedisGraph core developers was amazed by the results and said: ``\emph{I guess you are developing some query-generator that automatically compares results between different Cypher databases. If indeed, this is quite interesting! Do you plan to share it on GitHub?}'' He also said: ``\emph{We, at Redis, find the tool you are developing very interesting ...}'' One of Memgraph core developers replies with the following message for bugs reported by us: ``\emph{Aha, thanks for the reference ... I think Memgraph is not consistent and we should fix the behavior (that's a win-win). We are trying to follow openCypher as much as possible.}'' The feedback shows that our reported bugs are critical to the graph database engines. The feedback from real-world developers is also strong evidence that \GDsmith{} is practical for testing real-world graph database engines and able to detect critical bugs.

\subsection{Threats to Validity}

\subsubsection{External Validity}

The main threat to external validity is that the subjects chosen in our evaluations might not be generalized to other projects. To reduce the threat, we pick three well-known and open-source graph database engines as representatives. Their developers are active in open source communities and provide timely feedback on bug reports that we have filed. In fact, \GDsmith{} is able to test any graph database engine supporting the Cypher language.

\subsubsection{Internal Validity}

The main threat to internal validity lies in the implementation of \GDsmith{}. Not all Cypher features are currently supported (e.g., \CodeIn{UNION} clauses). To mitigate the threat, we investigate the covered Cypher language by multiple graph database engines, and refer to the description of core Cypher's syntax and semantics in previous work, enabling \GDsmith{} to support commonly used grammars.

\section{Related Work}
\label{sec:work}

There are two main categories of related work for \GDsmith{}.

\textbf{Detecting Bugs in Critical Software Systems.} Csmith~\cite{DBLP:conf/pldi/YangCER11} is a random C program generator for testing C compilers. It uses a sophisticated stochastical model to avoid generating C programs that have undefined behavior.  TVMfuzz~\cite{DBLP:conf/sigsoft/ShenM0TCC21} conducts mutation-based fuzzing to test deep learning compilers with some novel mutation operators to facilitate type-related bug detection. CYNTHIA~\cite{DBLP:conf/icse/SotiropoulosCAM21} detects bugs in object-relational mapping (ORM) implementations by employing a solver-based approach for generating targeted database records with respect to the constraints of the generated queries. To detect bugs in Datalog engines, queryFuzz~\cite{DBLP:conf/sigsoft/MansurCW21} uses metamorphic transformations based on database theory and performs metamorphic testing. OpFuzz~\cite{DBLP:journals/pacmpl/WintererZS20} and TypeFuzz~\cite{DBLP:journals/pacmpl/ParkWZS21} leverage different mutation strategies of formula generation to test satisfiability modulo theory (SMT) solvers. 
Compared with these approaches, \GDsmith{} makes the first attempt to address the challenges during automated test generation for graph database engines (which also belong to critical software systems), and detects 27 previously unknown bugs.

\textbf{Detecting Bugs in Relational Database Engines.} SQLsmith~\cite{SQLsmith} continuously generates syntactically correct SQL queries from the abstract syntax tree (AST) directly, meanwhile detecting whether the relational database engine under test faces  crashes. Squirrel~\cite{DBLP:conf/ccs/ZhongC0ZLW20} combines coverage-based fuzzing and model-based generation. It performs type-based mutations on the defined DSL and optimizes for semantic validity with additional analysis. 
Ratel~\cite{DBLP:conf/icse/WangWXLZZ021} is an enterprise-level fuzzer that improves the feedback precision, enhances the robustness of input generation, and performs an on-line investigation on the root cause of bugs with its industry-oriented design. 
The aforementioned approaches can detect only crashing bugs in relational database engines. To detect wrong-result bugs, 
RAGS~\cite{DBLP:conf/vldb/Slutz98} generates and executes SQL queries in multiple relational database engines, meanwhile observes differences in the output sets. Any inconsistency among results indicates at least one relational database engine contains bugs.
PQS~\cite{DBLP:conf/osdi/RiggerS20} detects wrong-result bugs by checking whether a specific record is fetched correctly. NoREC~\cite{DBLP:conf/sigsoft/RiggerS20} detects bugs in relational database engines by applying a semantics-preserving transformation to a given SQL query to disable the engine's optimizations and addresses PQS’ high implementation effort. TLP~\cite{DBLP:journals/pacmpl/RiggerS20} derives multiple SQL queries that compute a partial result of the initial query. By using a composition operator, the partitions can be combined to yield the same result as the original query; if the result differs, a bug in the relational database engine has been detected.
MutaSQL~\cite{DBLP:conf/sigmod/ChenWC20} generates test cases by mutating a SQL query over a database instance into a semantically equivalent query mutant, and checks the results returned by the relational database engine under test. 
Compared with these approaches, \GDsmith{} includes our skeleton-based completion technique to ensure that each randomly generated Cypher query satisfies the semantic requirements. \GDsmith{} also includes our novel techniques to increase the probability of producing Cypher queries returning non-empty results and improve the behavior diversity of generated queries, and these techniques are designed according to unique features of the Cypher language.

\section{Conclusion}
\label{sec:conclusion}

In this paper, we have introduced a new important problem of testing graph database engines. We have proposed \GDsmith{}, the first black-box testing approach for detecting bugs in graph database engines. We have implemented \GDsmith{} in Java and evaluated it against the baseline. The evaluation results demonstrate \GDsmith{}'s effectiveness and efficiency. We have also applied it to test three popular open-source graph database engines, successfully detecting \textbf{27 previously unknown bugs} on the released versions and receiving positive feedback from their developers.

In future work, we plan to improve our approach to support more features of the Cypher language and detect more bugs in real-world graph database engines. We also plan to conduct a comprehensive empirical study of bugs in graph computing systems and design an automated test generation approach for testing graph computing systems.


\bibliographystyle{ACM-Reference-Format}
\bibliography{sample-base}

\end{document}